\documentclass[namedreferences]{solarphysics}
\usepackage[optionalrh]{spr-sola-addons} 
\usepackage{graphicx}        
\usepackage{color}           
\usepackage{url}             
\usepackage[pdfborder={0 0 0 },urlcolor=blue,breaklinks]{hyperref}
\ifx \arxivurl  \undefined \def \arxivurl#1{\href{http://arxiv.org/abs/#1}{\textsf{arXiv}}}\fi 
\ifx \doiurl    \undefined \def \doiurl#1{\href{http://dx.doi.org/#1}{\textsf{DOI}}}\fi 
\ifx \adsurl    \undefined \def \adsurl#1{\href{http://adsabs.harvard.edu/abs/#1}{\textsf{ADS}}}\fi 

\begin{document}

\sloppy

\begin{article}

\begin{opening}

\title{Photospheric Magnetic Field: Relationship Between North--South Asymmetry and
Flux Imbalance\\ }

\author{E.S.~\surname{Vernova}$^{1}$\sep
        M.I.~\surname{Tyasto}$^{1}$\sep
        D.G.~\surname{Baranov}$^{2}$
       }
\runningauthor{E.S.~Vernova \textit{et al.}}
\runningtitle{North--South Asymmetry and Flux Imbalance}

   \institute{$^{1}$ IZMIRAN, SPb. Filial, St.~Petersburg, Russian Federation
                     email: \url{helena@ev13934.spb.edu} \\
              $^{2}$ A.F.~Ioffe Physical-Technical Institute,
                          St.~Petersburg, Russian Federation
                     email: \url{d.baranov@mail.ioffe.ru} \\
             }

\begin{abstract}
Photospheric magnetic fields were studied using the Kitt Peak synoptic
maps for $1976-2003$. Only strong magnetic fields ($B>100$\,G) of
the equatorial region were taken into account. The north--south
asymmetry of the magnetic fluxes was considered as well as the
imbalance between positive and negative fluxes. The
north--south asymmetry displays a regular alternation of the
dominant hemisphere during the solar cycle: the northern
hemisphere dominated in the ascending phase, the southern one in
the descending phase during Solar Cycles $21-23$. The sign of the
imbalance did not change during the 11 years from one polar-field
reversal to the next and always coincided with the sign of the
Sun's polar magnetic field in the  northern hemisphere. The
dominant sign of leading sunspots in one of the hemispheres
determines the sign of the magnetic-flux imbalance. The sign of
the north--south asymmetry of the magnetic fluxes and the sign of
the imbalance of the positive and the negative fluxes are related
to the quarter of the 22-year magnetic cycle where the magnetic
configuration of the Sun remains constant (from the minimum where
the sunspot sign changes according to Hale's law to the
magnetic-field reversal and from the reversal to the minimum). The
sign of the north--south asymmetry for the time interval
considered was determined by the phase of the 11-year cycle
(before or after the reversal); the sign of the imbalance of the
positive and the negative fluxes depends on both the phase of the
11-year cycle and on the parity of the solar cycle. The results
obtained demonstrate the connection of the magnetic fields in
active regions with the Sun's polar magnetic field in the northern
hemisphere.

\end{abstract}


\end{opening}

\section{Introduction}
\label{secintro} The principal cause of all manifestations of
solar activity is the magnetic field of the Sun, which, through
the process of its development, displays a surprising symmetry.
The magnetic field of the Sun displays the 22-year periodicity in
two ways (see, \textit{e.g.}, \opencite{charbonneau10}). The first
of the 22-year cycles is related to the global magnetic field of
the Sun (as a magnetic dipole) with a change of sign every 11
years, after the solar-activity maximum. Another 22-year
periodicity is connected with local magnetic fields (magnetic
fields of sunspots). According to the Hale polarity law, at each
minimum of solar activity, sunspots of a new cycle appear in each
of the Sun's hemispheres with the sign of the leading sunspots
opposite to the sign of the following sunspots in the same
hemisphere and opposite to the sign of the leading sunspots of
the other hemisphere. The signs of the leading and following
sunspots remain constant up to the following minimum when sunspots
of a new 11-year cycle appear with polarities opposite to the
previous ones. Thus, the magnetic field of the Sun varies with a
22-year periodicity that manifests itself both in the global
magnetic field change, and in the change of local magnetic fields
of the Sun.

There exist certain regularities in the changes of the global and
the local magnetic fields during a 22-year solar cycle. When one
considers the two magnetic cycles as a unique process, we see that
in the course of a 22-year magnetic cycle there are moments when
the relation between the polarities of the global and local
magnetic fields changes. These moments are related either to the
change of the Sun's global magnetic-field polarity during high
solar activity, or to the alternation of the polarities of leading
and following sunspots at the solar-activity minimum. Thus, during
a 22-year cycle there are four intervals where the polarities of
the global magnetic field and of the leading sunspots for each of
the hemispheres are constant. These four intervals are from the
solar-activity minimum to the magnetic-field reversal of the Sun
and from the reversal to the next minimum in an even 11-year solar
cycle, and analogously for the next odd cycle. The relations
between the polarities of the global and local magnetic fields
will repeat themselves in the next 22-year cycle. There is one
remarkable feature that we used in addition to this: for all
ascending phases of solar activity from a minimum to the reversal
the sign of the global field always coincides with the sign of the
leading sunspot in the same hemisphere: for the northern
hemisphere for an even cycle these signs will be minus, and for an
odd cycle the signs are plus.

Thus, seen on a longer time scale, the distributions of the local
and the global magnetic field exhibit a symmetry with respect to
the solar Equator. At the same time, an asymmetry is observed in
the magnetic fields, that manifests itself in three different
forms: as the longitudinal asymmetry, as the north--south
asymmetry, and as the asymmetry of the leading and following
sunspots in an active region.

We found it to be essential to divide  a 22-year magnetic cycle
into four intervals where the polarities of the local and global
magnetic fields are constant  when we considered the longitudinal
distribution of sunspots \cite{vernova04} and of photospheric
magnetic fields \cite{vernova07}. The longitudinal distribution
for the ascending phase and the maximum sharply differ from the
longitudinal distribution for the descending phase and the minimum
of the solar cycle. Active longitudes change by $180^\circ$ when
we pass from the ascending to the descending phase. The maximum of
the longitudinal distribution is reached at $180^\circ$ for the
ascending phase and the maximum of the solar cycle when the
polarities of the leading spot and of the global field coincide in
each hemisphere, and at $0^\circ/360^\circ$ for the descending
phase and the minimum when these polarities are opposite.

An analogous pattern of the active longitudes was discovered for
coronal mass ejections by \inlinecite{skirgiello05}. In a manner
similar to the results of \inlinecite{vernova04} and
\inlinecite{vernova07}, two nearly antipodal longitudes dominated
alternately: the domination of the longitude $180^\circ$ coincided
with the ascending phase and the maximum, while the domination of
the longitude $30^\circ$ coincided with the descending phase and
the minimum of the solar cycle.

There also exists an asymmetry between the leading and following
sunspots in an active region \cite{bray64, vitinskii86}.
\inlinecite{solanki03} in his survey considered the question: "Why
are leading polarity spots in an active region often larger than
following polarity spots?" as one of the important open questions.
Asymmetry of the leading and following polarities of an active
region are of interest and are discussed in many publications (see
\opencite{vandriel90}; \opencite{fan93}; \opencite{fan09}, and
references therein).

Another aspect of the solar-activity asymmetry is the north--south
asymmetry that was discovered when investigating various forms of
solar activity, such as sunspots, flares, or sudden disappearances
of solar prominences (see \opencite{carbonell07};
\opencite{verma09}; \opencite{sykora10}, and references therein).
It was observed that there are some common features in the
behavior of the north--south asymmetry for different
solar-activity indices \cite{vizoso90, li00}. For example, during
the period of the northern sunspot dominance, 78\,\% of solar
flares were observed in this hemisphere \cite{swinson86-2}.
However, it was shown by \inlinecite{roy77} that the north--south
asymmetry of solar flares does not follow the 11- or 22-year cycle
of the occurrence of major flares. The north--south asymmetry of
sunspot areas (see \opencite{ballester05}, and references therein)
shows three significant peaks in the power spectrum with periods
of 43.25, 8.65, and 1.44 years. For the north--south asymmetry of
sunspot groups the long-term period has been found to be about 80
years \cite{waldmeier57, pulkkinen99} and about 110 years
\cite{verma09}.

The north--south asymmetry is important for the topology of the
interplanetary space, and it influences both interplanetary and
near-Earth space parameters. For example, it affects the position
of the heliospheric current sheet (see \opencite{mursula03};
\opencite{wang11}, and references therein) and leads to a
difference in the number of away-from-the-Sun and toward-the-Sun
interplanetary-magnetic-field-sector days \cite{swinson91}. Since
some mechanisms of cosmic-ray modulation depend on the current
sheet, the north--south asymmetry of solar activity must be taken
into consideration when investigating cosmic-ray modulation.

Of special interest in the framework of the present article is the
change of the north--south asymmetry in the course of 11-year
solar cycle. \inlinecite{newton55} and \inlinecite{waldmeier71}
noted that the North dominates during the ascending phase of the
cycle while the South dominates during the declining phase.
\inlinecite{vizoso87} showed that for Cycles $18-21$ the sign of
the north--south asymmetry of the sudden disappearance of solar
prominences changed from positive to negative during the solar
maximum at the time of the Sun's magnetic-field reversal.
\inlinecite{yadav80} studied  the north--south asymmetry of solar
flares and showed that the majority of flares was concentrated in
the northern hemisphere before 1970 and in the southern hemisphere
after 1970 (the year of the reversal).

The heliographic distribution of significant ${\rm H}\alpha$ and
X-ray flares has been studied during the period $1975-1989$
\cite{curto09}, who found that the northern hemisphere dominates
in the ascending part of each solar cycle and the southern
hemisphere dominates in the descending part. \inlinecite{li00}
collected 20 different solar-activity phenomena to investigate
solar activity in Solar Cycle 21, and found that solar activity
has a northern bias at first, and as the cycle progresses,
southern predominance takes over. The pronounced north--south
asymmetry in the distribution and the number of CMEs in Cycle 23
observed by the \textit{Solar and Heliospheric Observatory} spacecraft was
reported by \inlinecite{minar08}: this asymmetry is in favor of
the northern hemisphere at the beginning of the cycle, and of the
southern hemisphere from 2001 (the year of the reversal) onward.

The behavior of the north--south asymmetry during the Sun's global
magnetic-field reversal is of special importance. The transition
of the dominant role from the northern hemisphere to the southern
one during the time of the reversal was found for sunspot area for
Cycles $16-22$ \cite{vernova02} and for the solar-flare index and
solar-group area for Solar Cycles $21-22$ \cite{kane05}.
\inlinecite{vernova02} studied the north--south asymmetry using a
vector summation of the sunspot areas that reduced the stochastic,
longitudinally evenly distributed sunspot activity and, therefore,
emphasized the more systematic, longitudinally asymmetric sunspot
activity. We found a systematic alternation of the dominant
hemisphere during high solar-activity periods, which is reproduced
from cycle to cycle. This effect was analyzed for Cycles $16-22$
by a superposed-epoch method using the date of magnetic reversal
in the southern hemisphere as the zero-epoch time to obtain an
average pattern of the north--south asymmetry. \inlinecite{kane05}
discovered analogous patterns (although with slightly different
lengths of the domination periods) for solar-flare index and
solar-group area for Solar Cycles 21 and 22.

The connection of north--south asymmetry with the phase difference
of solar activity in the two hemispheres was established in many
articles. \inlinecite{waldmeier71} has shown that because of a
phase shift, the hemisphere preceding in time is more active on
the ascending branch of the solar cycle, whereas on the descending
branch the hemisphere following in time dominates. The phase shift
changes its sign every four cycles so that the full period
contains eight 11-year cycles.  This periodicity was observed for
Solar Cycles $10-20$.

Various  methods were proposed to distinguish the two effects:
real asymmetry, \textit{i.e.} difference between the strength of
the solar cycle in the two hemispheres, and phase asymmetry. By
using the cross-recurrence plot technique it was shown that the
north--south sunspot asymmetry is due to phase asynchrony between
northern and southern hemispheric activities
(\opencite{ponyavin06}; \citeyear{ponyavin09}). The persistence of
phase-leading in one of the hemispheres exhibited a secular
variation. Beginning from Solar Cycle 13 and up to the end of
Cycle 16, the North was leading in time. After this epoch and
until the beginning of Cycle 20, the situation was reversed. After
Cycle 20 it was restored to the situation that was observed before
Cycle 16. Later these results were extended for the period of 300
years as far back as to the Maunder Minimum \cite{ponyavin10}.

\inlinecite{deng13} have analyzed the phase asynchrony of
hemispheric flare activity for the period of $1966-2008$. During
this period, flare activity in the northern hemisphere developed
six months earlier than that in the southern hemisphere. The
empirical-mode decomposition showed that the main periodicities of
flare activity in the northern hemisphere slightly differ from
those in the southern one, which should also lead to phase
asynchrony between them.

Another approach was used by \inlinecite{murakozy12}, who
represented the cycle profile by its centre of mass. For Solar
Cycles $12-23$ the phase of the hemispheric cycles shows an
alternating variation: the northern cycle led in four cycles and
followed in four cycles. This period approximately corresponded to
the length of the Gleissberg cycle. As pointed out by the authors,
this periodicity may not be absolutely rigid if the mechanism
controlling this variation is not part of the solar dynamo.
Although this variation persisted for the last 14 solar cycles, it
seems that phase relations are violated for Solar Cycle 24
\cite{svalgaard13} where the northern hemisphere again precedes
the southern one, contrary to the expected change of domination.

The magnetic fluxes of the Sun and their imbalance have been
studied by many authors on the basis of different data that
characterize the magnetic activity.  The problem of the nature of
the imbalance is widely discussed in the literature. As stated by
\inlinecite{kotov09}, the theory cannot explain why the positive
or the negative field dominates for one year or more practically
over the whole Sun, or, more precisely, over the visible
hemisphere. The observed prevalence of the negative mean magnetic
field of the Sun as a star in $1968-1969$ and $1989-1990$ could
not be compensated by the northern field of the polar regions.

\inlinecite{choudhary02} studied the magnetic-flux imbalance of
active regions using the longitudinal magnetograms obtained from
the National Solar Observatory at Kitt Peak. The median value of
the flux imbalance in 137 active regions was found to be about
9.5\,\%. The global-flux imbalance in the active latitude zone of
$10^\circ-40^\circ$ is solar-cycle dependent. The flux imbalance
is lower than 5\,\% when the active zones located in both
hemispheres are considered.

As demonstrated by these results, there are numerous
manifestations of asymmetric features of solar activity and the
Sun's magnetic fields. Further research is needed to combine all
varieties of these phenomena within the framework of the generalized
model of the solar dynamo.

The present article is concerned with the asymmetry of strong
photospheric magnetic fields ($B>100$\,G) of the equatorial region
(helio-latitudes from $-40^\circ$ to $+40^\circ$). We consider the
north--south asymmetry of the magnetic fluxes as well as the
imbalance between the positive and negative fluxes.

In Section~\ref{secdatameth} we describe the data and discuss the
method applied in the article. Section~\ref{secnsasym} is devoted
to the north--south asymmetry of the photospheric magnetic field.
In Section~\ref{secnands}, positive and negative magnetic fields
and their imbalance in each of the solar hemispheres are
considered. The imbalance of positive and negative magnetic fluxes
for the whole latitude range from $-40^\circ$ to $+40^\circ$ is
considered in Section~\ref{secimbal}, while in
Section~\ref{seclead} we are concerned with the relations between
the imbalance of the fluxes and the properties of the leading and
following sunspots. In Section~\ref{secconcl} we discuss and
interpret the  results obtained and draw our conclusions.

\section{Data and Method}
\label{secdatameth}  For this study  we used synoptic maps of the
photospheric magnetic field produced at the National Solar
Observatory/Kitt Peak  (available at
\href{http://nsokp.nso.edu}{nsokp.nso.edu}). These data cover the
period from 1975 to 2003 (Carrington Rotations $1625-2006$).
Because the data have many gaps during the initial period of
observations, we included in our analysis the data starting from
Carrington Rotation 1646. Synoptic maps have the following spatial
resolution: $1^\circ$ in longitude (360 steps); 180 equal steps in
the sine of the latitude from $-1$ (south pole) to $+1$ (north
pole). Thus every map consists of $360\times 180$ pixels of
magnetic-flux values.

Strong magnetic fields of both polarities occupy a relatively small
part of the Sun's surface. The magnetic-field strength for the period
$1976-2003$ shows a nearly symmetric distribution with 98.7\,\% of
values in the $0-100$\,G interval, whereas pixels with magnetic
strength above $100$\,G occupy only 1.3\,\% of the solar surface.
Even so, the total number of the latter pixels is large enough,
amounting to $3\times 10^5$ for $1976-2003$, and thus allowing
detailed analysis of strong magnetic fields and their temporal
changes.

We focused here on the photospheric magnetic fields connected with
solar active regions. To do this, we limited our research to the
near-equatorial and strong magnetic fields. From here on only
equatorial magnetic fields are considered, \textit{i.e.}
helio-latitudes from $-40^\circ$ to $+40^\circ$. To calculate the
magnetic flux only pixels with magnetic field strengths higher
than $100$\,G  were selected. This approach allowed us to
investigate changes of the photospheric magnetic field connected
with active areas on the Sun, among which bipolar sunspot groups
play an important role. For each synoptic map, magnetic fluxes for
the northern and southern hemispheres were analyzed separately.
Fluxes for positive and negative magnetic fields were calculated
for each of the hemispheres. Thus, for each synoptic map four
different characteristics of magnetic flux were obtained: $F_{{\rm
N}}^{{\rm pos}}$, $F_{{\rm N}}^{{\rm neg}}$, $F_{{\rm S}}^{{\rm
pos}}$, $F_{{\rm S}}^{{\rm neg}}$. For each solar cycle, positive
and negative fluxes in the northern hemisphere [$F_{{\rm N}}^{{\rm
pos}}$ and $F_{{\rm N}}^{{\rm neg}}$] coincide in sign with the
fields of the leading or following sunspots. The same is true for
the southern hemisphere where the polarities of the leading
(following) sunspots are opposite to those of the northern
hemisphere. It is expected that under our restrictions (strong
magnetic fields of the sunspot zone) each of the four fluxes is
connected with the sunspots of either leading or following
polarity, but they are not identical to the sunspot fluxes. We
studied the changes of these fluxes, as well as temporal changes
of the difference between positive and negative fluxes.

\section{North--South Asymmetry of the Photospheric Magnetic Field}
\label{secnsasym} We considered the total flux of the northern
hemisphere (latitudes from $0^\circ$ to $+40^\circ$) as the sum of
absolute values both of positive and negative fluxes [$F_{{\rm N}}
= |F_{{\rm N}}^{{\rm pos}}| + |F_{{\rm N}}^{{\rm neg}}|$] and the
total flux of the southern hemisphere (latitudes from $0^\circ$ to
$-40^\circ$) as $F_{{\rm S}}  =  |F_{{\rm S}}^{{\rm pos}}|
+|F_{{\rm S}}^{{\rm neg}}|$, respectively.

\begin{figure}
\begin{center}
\includegraphics[width=0.75\textwidth]{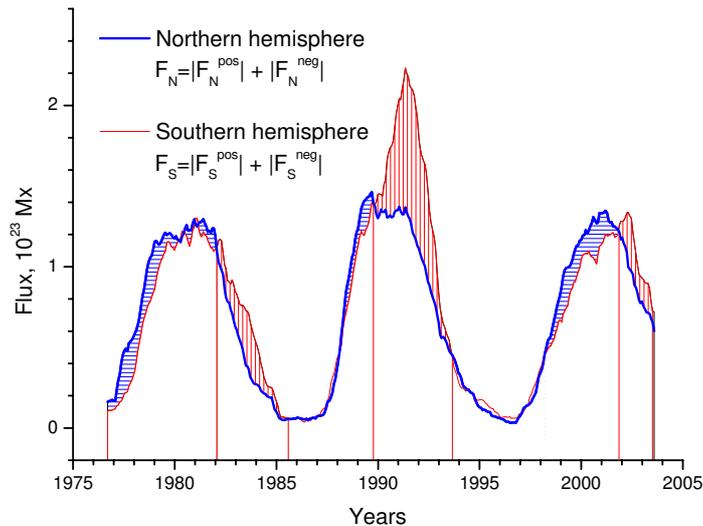}
\caption{Photospheric magnetic flux: northern hemisphere [$F_{{\rm N}}  =
|F_{{\rm N}}^{{\rm pos}}|  + |F_{{\rm N}}^{{\rm neg}}|$] -- thick blue curve; southern hemisphere
[$F_{{\rm S}}  = |F_{{\rm S}}^{{\rm pos}}| +|F_{{\rm S}}^{{\rm neg}}|$] -- red curve. Only strong
magnetic fields [$B>100$\,G] of the equatorial region
(helio-latitudes from $-40^\circ$ to $+40^\circ$) were taken into
account. Periods of northern hemisphere domination are marked
by horizontal blue shading. Vertical red shading marks southern hemisphere
domination. Vertical lines show the change of the dominant
hemisphere. }
\label{nshemisph}       
\end{center}
\end{figure}

In  Figure~\ref{nshemisph}, total magnetic fluxes for the northern
and southern hemispheres are shown for 1976\,--\,2003. To exclude
random fluctuations, flux values were smoothed using adjacent
averaging over 20 solar rotations. Both fluxes change following
the 11-year cycle of solar activity. However, some differences
between  fluxes of the northern and southern hemispheres can be
seen. During the ascending  phase of the solar cycle, the northern
hemisphere dominates, while during the descending phase the
southern hemisphere always dominates the northern one. These
features occur regularly throughout Solar Cycles $21-23$.

\begin{figure}
\begin{center}
\includegraphics[width=0.75\textwidth]{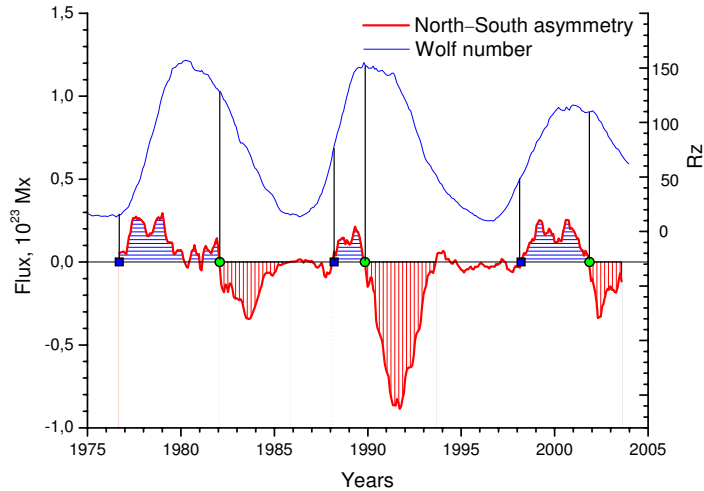}
\caption{north--south asymmetry of the photospheric magnetic flux
(thick red
   line):  $\Delta F_{{\rm NS}} = (|F_{{\rm N}}^{{\rm pos}}| + |F_{{\rm N}}^{{\rm neg}}|) - (|F_{{\rm S}}^{{\rm pos}}| + |F_{{\rm S}}^{{\rm neg}}|$).
   Periods of northern hemisphere domination are shown by horizontal blue shading;
   vertical red shading corresponds to southern hemisphere
   domination.  Vertical lines show the change of the dominant
   hemisphere. Wolf numbers [Rz] (blue line) are plotted for comparison of
   magnetic flux changes with the solar-activity cycle. The change of
   the north--south asymmetry sign is marked by  squares (ascending phase of the
   solar cycle) and by circles (period of the global magnetic field
   reversal). } \label{nsasym}
\end{center}
\end{figure}

This result becomes  more evident if one considers the
difference of fluxes
\begin{equation}
\label{deltans} \Delta F_{{\rm NS}} = (|F_{{\rm N}}^{{\rm pos}}| + |F_{{\rm N}}^{{\rm neg}}|) -
(|F_{{\rm S}}^{{\rm pos}}| + |F_{{\rm S}}^{{\rm neg}}|),
\end{equation}
that is the north--south asymmetry of the magnetic field
(Figure~\ref{nsasym}). The difference [$\Delta F_{{\rm NS}}$] is
close to zero during periods of low solar activity. In each of the
solar cycles, the difference passes twice through zero: during the
global magnetic-field reversal and during the solar minimum. As a
result, we see a regular change of alternately dominating
hemispheres. In the ascending phase (from the solar-activity
minimum to the reversal) the northern hemisphere always dominates
(the difference has positive sign). During reversal, the
difference becomes negative, \textit{i.e.} the leading role passes
to the southern hemisphere. Domination of the southern hemisphere
continues from the reversal to the next minimum. Near the
solar-activity minimum the sign of the difference changes again
and the northern hemisphere becomes dominant.

Thus, we draw the following conclusions: we have considered the
north--south asymmetry of the photospheric magnetic field for
almost three solar cycles (Solar Cycles $21-23$) and for each of
them we have observed the same rule of the north--south asymmetry
change in the 11-year solar cycle. This rule can be formulated as
follows: each of the 11-year cycles can be divided into two parts,
the ascending phase and the descending phase of solar activity.
For each of these parts, the following effects take place:

i) In the ascending phase the northern hemisphere always dominates
the southern one;

ii) During reversal the dominant hemisphere
changes so that in the descending phase to the solar minimum
the southern hemisphere always dominates.

This effect cannot be seen in the  north--south asymmetry averaged
for long-term periods such as the 11-year cycle or longer. On the
other hand, strong fluctuations mask the effect if one considers
short time intervals. Yet these  features are clearly seen after
averaging of the data over 20 solar rotations. Maxima of the
north--south asymmetry amount to $\approx 30$\,\% of the average
magnetic flux for $1976-2003$. The close connection of the
asymmetry with the solar-cycle phase is clearly seen during the
three cycles.

We emphasized in the Introduction that characteristic points of
the 22-year magnetic cycle play important roles for the
solar-activity development. These points are: i) reversal of the
global magnetic field of the Sun (after the solar-activity
maximum) and ii) change of sign of the leading and following
sunspots in each of the hemispheres according to Hale's law (at
solar-activity minimum). Between these points the sign of the
global magnetic field and the sign of leading sunspots for each of
the hemispheres is constant. Thus, for each quarter of the 22-year
magnetic cycle the polarities of the global and local magnetic
fields of the Sun remain constant.

As was shown, in the ascending phase up to the reversal, the
magnetic flux of the northern hemisphere  always dominates when
compared with the flux of the southern hemisphere. At this time
the leading sunspots of the northern hemisphere have the same
polarity as the global field in the northern hemisphere. After the
reversal, the sign of the global field in the northern hemisphere
coincides with the sign of leading sunspots in the southern
hemisphere. At the same time, we see that the dominant
hemisphere has changed as well: domination passes to the southern
hemisphere.

An important conclusion follows: leading sunspots of the dominant
hemisphere have the same polarity as the global magnetic field in
the northern hemisphere. It should be noted that these results are
true for the period of Solar Cycles $21-23$ when the activity of
the northern hemisphere preceded that of the southern hemisphere
(see Introduction).

These results are summarized in Table~\ref{polarity}. We present
the 22-year period (Cycles 21 and 22), divided into four quarters.
Commonly, two consecutive solar cycles are combined into a 22-year
magnetic cycle,  starting with an even one \cite{gnevyshev48}.
Because we have no data for the whole of Cycle 23, we considered
the odd--even pair of cycles. The first quarter of the 22-year
period includes the ascending phase of Cycle 21 from the minimum
to the reversal; the second quarter is a period of the descending
phase of Cycle 21 from the reversal to the minimum. Similarly,
Cycle 22 is divided into two parts. For each of these quarters the
following characteristics are shown: i) the global magnetic-field
sign (northern hemisphere); ii) the dominant hemisphere;
iii) the sign of leading sunspots of the dominant hemisphere. The
regular change of the dominant hemispheres can be seen in
Table~\ref{polarity}, as well as the coincidence of the global
magnetic-field sign in the northern hemisphere with the sign  of
the leading sunspots of a dominant hemisphere  for each of the
quarters.
\begin{table}
\caption{Polarity of solar magnetic fields and north--south
asymmetry for Solar Cycles $21-22$}
\label{polarity}       
\begin{tabular}{lcccc}
\hline\noalign{\smallskip} Solar Cycle & 21 & 21 & 22 &  22 \\
\noalign{\smallskip}  & from minimum & from reversal & from minimum & from reversal \\
\noalign{\smallskip}  & to reversal & to minimum & to reversal & to minimum \\
\noalign{\smallskip}\hline\noalign{\smallskip}
Quarter of the  & 1976\,--\,1981 & 1982\,--\,1986 & 1987\,--\,1991 & 1992\,--\,1996  \\
\noalign{\smallskip} 22-year period & I & II & III & IV \\
\hline
Global magnetic  &  &  &  &   \\
field sign:  & $+$ & $-$ & $-$ & $+$  \\
N hemisphere &  &  &  &   \\
\hline
Dominant &  &  &  &   \\
hemisphere & N & S & N & S  \\
\hline
Leading sunspot:  &  &  &  &   \\
dominant & $+$ & $-$ & $-$ & $+$  \\
hemisphere &  &  &  &   \\
\hline
Sign of the flux &  &  &  &   \\
imbalance & $+$ & $-$ & $-$ & $+$ \\
\noalign{\smallskip}\hline
\end{tabular}
\end{table}
\par

These features of the north--south asymmetry of the magnetic field
can be expressed by the following equation that establishes the
law of the north--south asymmetry sign change for the two parts of
the 11-year cycle -- before the global magnetic-field reversal and
after it:
\begin{equation}
\label{sign} {\rm sign}\,\Delta F_{{\rm NS}}  = (-1)^{{\rm k}+1}, \quad
\mbox{where}\ {\rm k}=1,2
\end{equation}
(${\rm k}=1$ corresponds to the interval from the minimum of the
11-year cycle to the reversal; ${\rm k}=2$ corresponds to the
interval from the reversal to the minimum). This formula shows
that from the minimum to the reversal the northern hemisphere
always dominates, while from the reversal to the minimum the
dominant hemisphere is the southern one. We stress that this
relation is obtained for Solar Cycles $21-23$ when the northern
hemisphere preceded the southern one in time (see Introduction).

For the three cycles that we considered, Equation~(\ref{sign})
correctly describes the change of the north--south asymmetry of
magnetic fields shown in Figure~\ref{nsasym}. As has been stated
in the Introduction, a number of works established the change of
the dominant hemisphere for the ascending and descending phases of
the solar cycle. This effect was observed for different
manifestations of solar activity (see \opencite{vizoso87}, and
references in the Introduction). A rapid transition from the
domination of the northern hemisphere to the domination of the
southern one after the magnetic-field reversal has been observed
for sunspot areas \cite{vernova02} and for solar flares
\cite{kane05}. A reverse change from the southern hemisphere to
the northern one during the solar minimum was found for sunspots
\cite{swinson86}. Because the change of the photospheric magnetic
field determines the cyclic change of various indices of solar
activity, the proposed formula can be interpreted as a
generalization of some of the results cited above.


\section{Positive and Negative Magnetic Fields and their Imbalance
for Each Hemisphere} \label{secnands} In the previous section,
absolute values of magnetic field were used to evaluate the total
magnetic flux. Now we consider the positive and negative fluxes of
the photospheric magnetic field separately to study their
imbalance for each of the solar hemispheres. As above, we only
take into account the strong magnetic fields ($B>100$\,G) of the
equatorial region (from $-40^\circ$ to $+40^\circ$).
\begin{figure}
\begin{center}
   \includegraphics[width=0.75\textwidth]{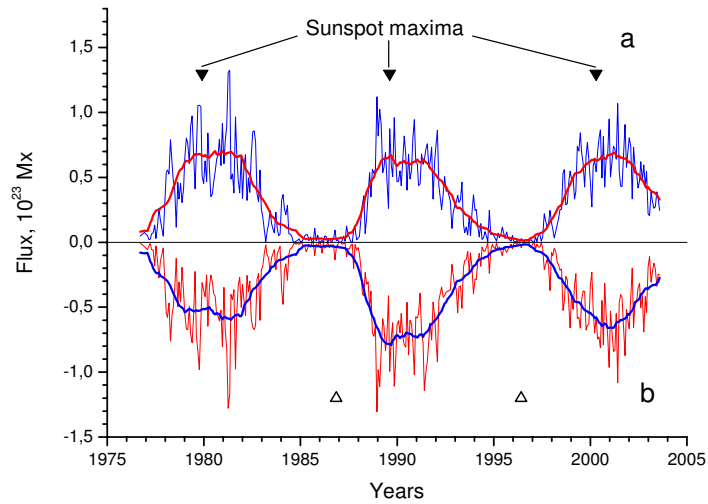}
   \caption{Fluxes of the strong magnetic fields [$B>100$\,G] of the
   northern hemisphere (helio-latitudes from $0^\circ$ to $+40^\circ$):
   positive (a) and  negative (b) photospheric magnetic fluxes. Thick lines
   represent data averaged over 20 solar rotations. Triangles mark
   solar maxima (black triangles) and solar minima (white triangles).
 }
\label{nposneg}       
\end{center}
\end{figure}
In Figure~\ref{nposneg} positive ($F_{{\rm N}}^{{\rm pos}} > 0$,
Figure~\ref{nposneg}a) and negative ($F_{{\rm N}}^{{\rm neg}} <0$,
Figure~\ref{nposneg}b) magnetic fluxes for the northern hemisphere
of the Sun are presented. Smoothed curves are shown, which were
obtained using adjacent averaging over 20 rotations. It can be
seen that both fluxes show a distinct 11-year periodicity. This is
expected because we consider the strong fields, which are known to
be connected with active regions. As a result, the magnetic flux
drops almost to zero during minima of solar activity.

\begin{figure}
\begin{center}
\includegraphics[width=0.75\textwidth]{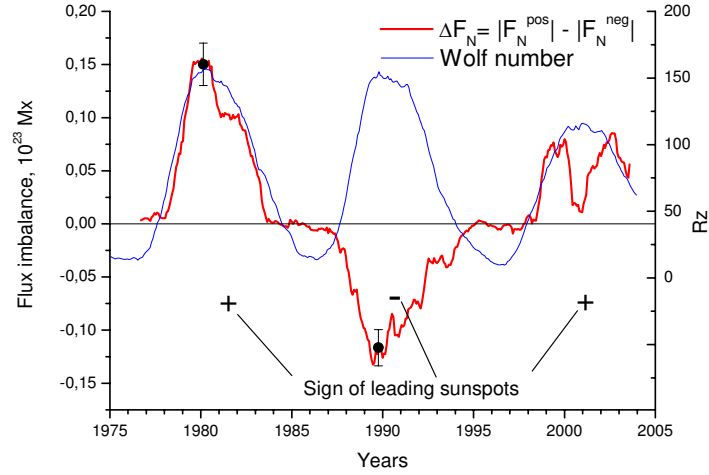}
   \caption{Positive and negative flux imbalance [$\Delta F_{{\rm N}} = |F_{{\rm N}}^{{\rm pos}}|
   - |F_{{\rm N}}^{{\rm neg}}|$] (northern hemisphere) -- thick red line.
   Wolf numbers [Rz] are plotted by the
   blue line. Sign of leading sunspots in the northern hemisphere for
   each of the three solar cycles is shown. Statistical spread of averaged
   data is given for extrema of the imbalance curve.}
\label{imbnorth2}       
\end{center}
\end{figure}

Positive and negative fluxes are well correlated, but their
difference [$\Delta F_{{\rm N}}= |F_{{\rm N}}^{{\rm pos}}| -
|F_{{\rm N}}^{{\rm neg}}|$] (flux imbalance or net flux) shows
regular changes closely connected with  the phase of the 22-year
solar cycle.  The smoothed value of the difference [$\Delta
F_{{\rm N}}$] for the northern hemisphere is shown in
Figure~\ref{imbnorth2}. Maxima of the difference amount to
$\approx 20$\,\% of the average magnetic flux for the northern
hemisphere. Wolf numbers are shown for comparison with the
solar-cycle progression. The difference varies with the 22-year
cycle and reaches extrema during maxima of solar activity. The
difference is close to zero for several years around the minima of
solar activity ($1984-1987$ and $1995-1998$). During these periods
the difference passes through zero. Thus, from one minimum to
another, the sign of $\Delta F_{{\rm N}}$ (the difference  between
positive and negative fluxes) does not change. In
Figure~\ref{imbnorth2} the sign of leading sunspots in bipolar
sunspot groups is shown. The sign of the difference coincides with
the sign of leading sunspots in the northern hemisphere.
Evidently, the leading sunspots contribute more to the magnetic
flux than the following ones.

\begin{figure}
\begin{center}
\includegraphics[width=0.75\textwidth]{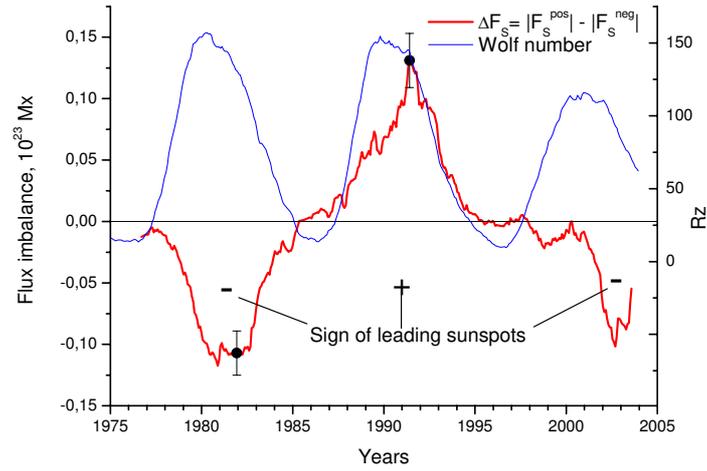}
   \caption{Positive and negative flux imbalance [$\Delta F_{{\rm S}} = |F_{{\rm S}}^{{\rm pos}}|
    - |F_{{\rm S}}^{{\rm neg}}|$] (southern hemisphere) -- thick red line. Wolf numbers [Rz]
   are plotted by the
   blue line. Sign of leading sunspots in the southern hemisphere for
   each of the three solar cycles is shown. Statistical spread of averaged
   data is given for extrema of the imbalance curve.}
\label{imbsouth2}       
\end{center}
\end{figure}
Similar results are obtained for the southern hemisphere: the sign
of the imbalance always coincides with the sign of leading
sunspots (Figure~\ref{imbsouth2}). Thus, a positive difference
corresponds to the time when leading sunspots in the given
hemisphere have positive polarity. In the same way, the negative
difference corresponds to the period when leading sunspots have
negative polarity. This effect can be observed for almost three
solar cycles both for the northern and southern hemispheres of the
Sun. The imbalance of positive and negative fluxes in each of the
solar hemispheres shows a 22-year recurrence that is directly
connected with the Hale cycle. The absolute values of the
imbalance for both solar hemispheres [$|\Delta F_{{\rm
N}}|,|\Delta F_{{\rm S}}|$] closely follow the sunspot activity
for Cycles 21 and 22 (Figures~\ref{imbnorth2}, \ref{imbsouth2}).
The correlation becomes somewhat poorer for Cycle 23, although the
main effect can be observed in this case as well: the change of
the imbalance sign during solar minimum.

These results agree with direct measurements of the magnetic field
of sunspots, which give a higher magnetic flux  for  the leading
sunspots than for the following ones \cite{vitinskii86,
choudhary02,fan09}. It has been shown that the balance of magnetic
fluxes is not observed for all classes of sunspot groups
\cite{vitinskii86}. Moreover, the longitude-averaged magnetic
field of active regions does not become zero, the sign of the
field being determined by the polarity of leading sunspots
\cite{obridko07}. For the northern hemisphere during odd solar
cycles, magnetic fields with positive polarity slightly exceed in
strength the fields of the opposite sign, occupying at the same
time smaller areas. Note that according to our results not only
the strength [$B$], but also the flux of positive fields in the
northern hemisphere  ($F=BS$, where $S$ is the sunspot area)
exceeds the flux of the opposite sign during odd cycles.

The imbalance between magnetic fields of different polarities
(Figures~\ref{imbnorth2} and \ref{imbsouth2}) may be connected
with  the asymmetry in morphology of the leading and following
sunspots. It is well known that the leading polarity of an active
region tends to be in the form of large sunspots, whereas the
following polarity tends to appear more dispersed and fragmented.
Numerical simulations of \inlinecite{fan93} showed that the area
asymmetry, as well as the field-strength asymmetry, between the
preceding and following polarities might be produced by the
Coriolis force during a flux tube's rising motion in the solar
convection zone. The area asymmetry of bipolar magnetic fields was
studied quantitatively by \inlinecite{yamamoto12} using
magnetograms of bipolar regions.  Areas of the preceding
polarities proved to be smaller than those of the following
polarities in many bipolar magnetic regions. Only active regions
with magnetic imbalances lower than 15\,\% were selected for this
study.  Several authors have reported higher magnetic-flux
imbalances in active regions (see \opencite{tian03}, and
references therein). About 47\,\% of the active regions are in a
10\,\% flux imbalance and 31\,\% of the active regions are in a
flux imbalance between 10\,\% and 20\,\%. Only 22\,\% of the
active regions have a flux imbalance higher than 20\,\%
\cite{tian03}. The lower limit of the field strength considered
was set at 50\,G by \inlinecite{yamamoto12} and at 20\,G by
\inlinecite{tian03}. In our study only strong magnetic fields
($B>100$\,G) were considered. One could argue that the
magnetic-field imbalance of the leading and the following polarity
that we observed (Figures~\ref{imbnorth2} and \ref{imbsouth2})
arose because weaker fields of the following sunspots were not
included in our calculation.

\begin{figure}
\begin{center}
\includegraphics[width=4in]{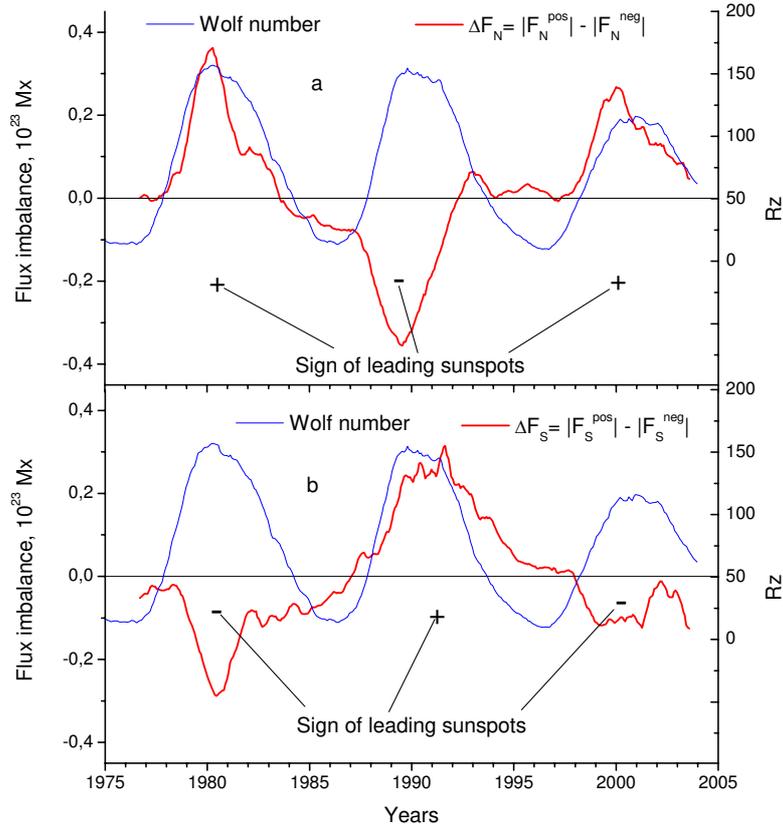}
   \caption{Positive and negative flux imbalance calculated for all values of
   magnetic-field strength (thick red line):
   (a) northern hemisphere [$\Delta F_{{\rm N}} = |F_{{\rm N}}^{{\rm pos}}|   - |F_{{\rm N}}^{{\rm neg}}|$];
   (b) southern hemisphere [$\Delta F_{{\rm S}} = |F_{{\rm S}}^{{\rm pos}}|   - |F_{{\rm S}}^{{\rm neg}}|$].
   Wolf numbers [Rz] are plotted by the  blue line. Sign of leading sunspots
   is shown for each of the three solar cycles.}
\label{imbnsall}       
\end{center}
\end{figure}

To check this possibility we extended our study of imbalance to
all values of magnetic-field strength (Figure~\ref{imbnsall}).

The imbalance of positive and negative fluxes proved to be even
higher than for strong magnetic fields (Figures~\ref{imbnorth2}
and \ref{imbsouth2}). This is true both for the northern and
southern hemispheres.  As pointed out by \inlinecite{svalgaard13},
magnetic flux from decaying sunspots moves toward the poles,
predominantly from the following spots. This may lead to a deficit
of the following polarity in the sunspot zone and produce the
observed imbalance.


\section{Imbalance of Positive and Negative Magnetic Fluxes for
the Whole Latitude Range from $-40^\circ$ to $+40^\circ$}
\label{secimbal}

Now we consider the imbalance of positive and negative magnetic
fluxes for the whole latitude interval from $-40^\circ$ to
$+40^\circ$. As earlier, we consider only strong magnetic fields
[$B>100$\,G].

\begin{figure}
\begin{center}
\includegraphics[width=0.75\textwidth]{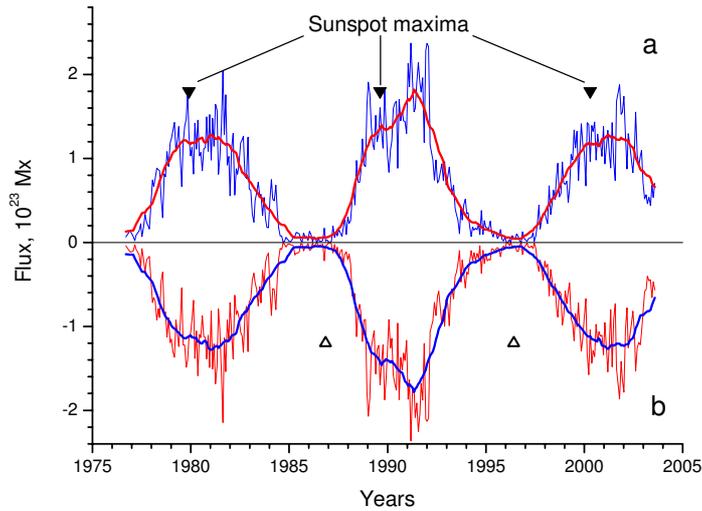}
\caption{Positive and negative fluxes of the strong magnetic
fields [$B>100$\,G] for both solar hemispheres together
(helio-latitudes from  $-40^\circ$ to $+40^\circ$): positive (a)
and negative (b) photospheric magnetic fluxes. Thick lines
represent data averaged over 20 solar rotations. Triangles mark
solar maxima (black triangles) and solar minima (white
triangles).}
\label{posneg}       
\end{center}
\end{figure}
\begin{figure}
\begin{center}
\includegraphics[width=0.75\textwidth]{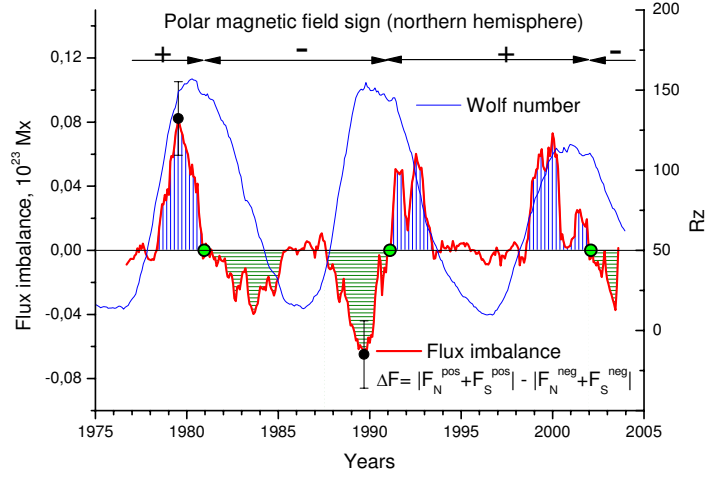}
   \caption{Positive and negative flux imbalance for both solar
   hemispheres $\Delta F = |F_{{\rm N}}^{{\rm pos}} + F_{{\rm S}}^{{\rm pos}}| - |F_{{\rm N}}^{{\rm neg}} + F_{{\rm S}}^{{\rm neg}}|$
   -- thick red line (helio-latitudes from $-40^\circ$ to $+40^\circ$).
   Periods of the positive polarity domination are shown by the vertical blue shading;
   horizontal green shading corresponds to the negative polarity domination. Green
   circles mark the change of the imbalance sign. Statistical spread
   of averaged data is given for extrema of the imbalance curve.
   Wolf numbers [Rz] are plotted by the blue line. Polar magnetic-field
   sign is shown for the northern hemisphere. }
\label{netflux2}
\end{center}
\end{figure}

Positive [$F_{{\rm N}}^{{\rm pos}}+ F_{{\rm S}}^{{\rm pos}}$] and
negative [$F_{{\rm N}}^{{\rm neg}} + F_{{\rm S}}^{{\rm neg}}$]
fluxes are presented in Figure~\ref{posneg}. Changes of positive
and negative fluxes are very similar, both follow the 11-year
solar cycle. However, the difference of these fluxes,
\textit{i.e.} the imbalance of positive and negative fluxes shows
regular changes in the course of a 22-year solar cycle (see our
results in \opencite{bulgaria09}). The difference between positive
and negative fluxes [$\Delta F$] is displayed in
Figure~\ref{netflux2}, where
\begin{equation}
\label{deltaf}
\Delta F = |F_{{\rm N}}^{{\rm pos}} + F_{{\rm S}}^{{\rm pos}}| - |F_{{\rm N}}^{{\rm neg}} + F_{{\rm S}}^{{\rm neg}}|.
\end{equation}

The difference was calculated for each of the solar rotations, and
then smoothed using adjacent averaging over 20 rotations. The
smoothed value of the difference is shown in Figure~\ref{netflux2}
in comparison  with the Wolf numbers. The difference between the
fluxes depends on the level of solar activity. The difference
becomes nonzero during periods of high solar activity and is
almost equal to zero around minima of solar activity ($1985-1988$
and $1994-1998$).

In the upper part of Figure~\ref{netflux2}, the polarity of the
polar magnetic field in the northern hemisphere is displayed.
Green circles mark the change of the sign of the imbalance. The
imbalance changes its sign near the global magnetic field
reversals. Thus, the sign of the dominant flux remains constant
within 11 years  from one global magnetic-field reversal to the
next. From the minimum of 1976 to the reversal of Cycle 21, the
imbalance is positive. During the reversal of the global magnetic
field (1981) it becomes negative and keeps this sign until the
next reversal (1991). Then the sign of the imbalance changes and
again becomes positive. The maximum of the imbalance amounts to
4\,\% of the average magnetic flux.

As Figure~\ref{netflux2} shows, the difference of fluxes changes
with the 22-year period, passing through  zero near to the global
magnetic-field reversal. Thus, the difference of fluxes preserves
its sign within 11 years from one reversal to another. The sign of
the difference always coincides with the polar magnetic-field sign
in the northern hemisphere. Before the reversal, the sign of the
difference coincides not only with the polar magnetic-field sign
in the northern hemisphere, but also with the  sign of leading
sunspots in the northern hemisphere. In contrast, after the
reversal, the sign of the difference coincides with the polar
magnetic-field sign in the northern hemisphere and with the sign
of leading sunspots in the southern hemisphere.

The sign of the imbalance for each of the four quarters of the
22-year magnetic cycle is shown in Table~\ref{polarity} compared
with the change of magnetic-field polarities.

The change of the sign of the positive and negative magnetic flux imbalance
$\Delta F$ can be expressed by the following formula:
\begin{equation}
\label{signf}
 {\rm sign}\,\Delta F = (-1)^{{\rm n}+{\rm k}},
\end{equation}
where  ${\rm n}=1,2$ (${\rm n}=1$ corresponds to the odd solar
cycle, ${\rm n}=2$ to the even one); ${\rm k}=1,2$ (${\rm k}=1$
corresponds to the interval of the 11-year cycle from the minimum
up to the reversal; ${\rm k}=2$ to the interval from the reversal
up to the minimum). In this way, the imbalance sign is determined
by two factors: the parity of the solar cycle on one hand, and the
phase of the 11-year cycle on the other hand.


\section{Imbalance of the Fluxes and the Asymmetry Between Leading
and Following Sunspots}
  \label{seclead}
We have seen that the sign of the imbalance of positive and
negative fluxes remains constant for 11 years from one reversal
to another and coincides with the sign of the polar magnetic
field of the Sun in the northern hemisphere.

The question arises whether it is possible to specify one of the
four magnetic fluxes [$F_{{\rm N}}^{{\rm pos}}$, $F_{{\rm
N}}^{{\rm neg}}$, $F_{{\rm S}}^{{\rm pos}}$, $F_{{\rm S}}^{{\rm
neg}}$] that plays the dominant role and determines the sign. When
we considered the imbalance of positive and negative flux in the
latitude range from $-40^\circ$  to $+40^\circ$ we took into
account the flux of the leading and following sunspots. Thus, the
magnetic flux of positive polarity contains the contribution of
the leading sunspots of one hemisphere, and the contribution of
the following sunspots of the opposite hemisphere.

\begin{figure}
\begin{center}
\includegraphics[width=0.75\textwidth]{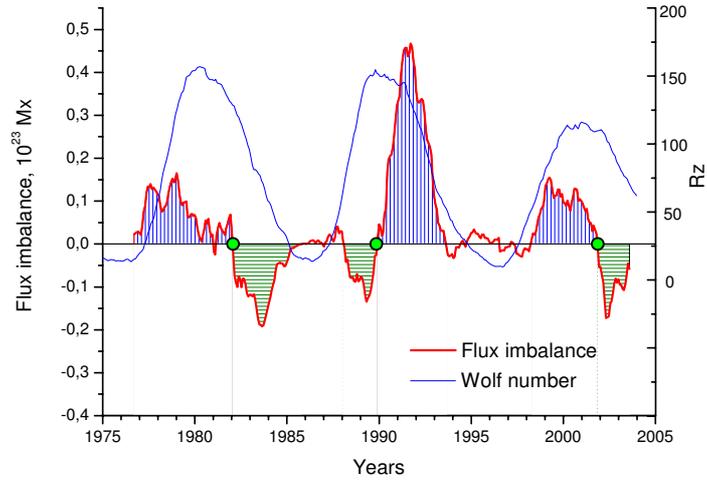}
\caption{Positive and negative flux imbalance of leading  sunspots for both solar
   hemispheres [$\Delta F_{{\rm L}} = |F_{{\rm N}}^{{\rm pos}}| - |F_{{\rm S}}^{{\rm neg}}|$]
   -- thick red line (helio-latitudes from $-40^\circ$ to $+40^\circ$).
   Periods of the positive polarity domination are shown by the vertical blue shading;
   horizontal green shading corresponds to the negative polarity domination. Green
   circles mark the change of the imbalance sign.
   Wolf numbers [Rz] are plotted by the blue line. }
\label{leading}
\end{center}
\end{figure}

Because for each solar cycle the sign of leading sunspots in each
of the Sun's hemispheres is known, it is possible to evaluate the
imbalance of the magnetic fluxes for the leading sunspots alone.
For example, for Solar Cycle 21 leading sunspots in the northern
hemisphere have a positive polarity [$F_{{\rm N}}^{{\rm pos}} >
0$], while leading sunspots of the southern hemisphere have a
negative sign [$F_{{\rm S}}^{{\rm neg}} <0$]. This means that the
difference [$\Delta F_{{\rm L}} = |F_{{\rm N}}^{{\rm pos}}| -
|F_{{\rm S}}^{{\rm neg}}|$] represents the imbalance of the
magnetic flux of leading sunspots during Cycle 21. An analogous
approach can be used for other solar cycles. The resulting
imbalance of the leading sunspot fluxes is shown in
Figure~\ref{leading}.

All of the basic features of the total imbalance
(Figure~\ref{netflux2}) are also observed in the imbalance of the
leading sunspot fluxes (Figure~\ref{leading}). The sign of the
dominant flux changes during the global magnetic-field reversal
and then remains constant for 11 years until the next reversal.
Similar to the net flux (Figure~\ref{netflux2}), the imbalance of
the leading sunspot fluxes (Figure~\ref{leading})is positive from
the solar minimum to the reversal of Cycle 21; around the time of
the reversal of the global magnetic field it becomes negative and
retains this sign until the next reversal (Figure~\ref{leading}).
Then the sign of the imbalance changes and again becomes positive.

\begin{figure}
\begin{center}
\includegraphics[width=0.75\textwidth]{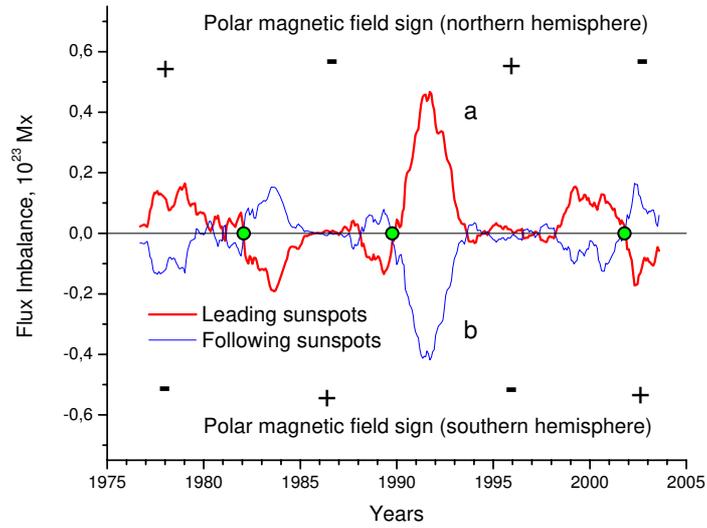}
   \caption{Flux imbalance of leading (thick red line) and following (blue
   line) sunspots of both hemispheres.
   Polar magnetic-field sign for the northern (respectively, southern)
   hemisphere is shown in the upper (respectively, lower) part of the figure.}
\label{leadfollow2}
\end{center}
\end{figure}

It should be noted that the imbalance of leading sunspot flux is
significantly higher than the total imbalance that includes all
four fluxes [$F_{{\rm N}}^{{\rm pos}}$, $F_{{\rm N}}^{{\rm neg}}$,
$F_{{\rm S}}^{{\rm pos}}$, $F_{{\rm S}}^{{\rm neg}}$]. The maxima
of the leading-sunspot imbalance amount to $\approx 15$\,\% of the
average magnetic flux. This effect occurs because the imbalance of
the following sunspots is in antiphase with the imbalance of
leading sunspots (Figure~\ref{leadfollow2}). Moreover, the
imbalance of leading and of following sunspots is related by an
almost linear dependence with the correlation coefficient
$R=-0.98$. This means that when the leading sunspots of one hemisphere
dominate over the leading sunspots of the other hemisphere,
the following sunspots in the first hemisphere dominate over
the following sunspots of the other hemisphere.

The imbalance between flux of leading sunspots in the two
hemispheres maintains its sign  for each quarter of a 22-year
magnetic cycle. The same is true for the following sunspots, so
that the northern hemisphere dominates from a minimum to a
reversal, and the southern hemisphere dominates from a reversal to
a minimum for both leading and following sunspots. Simultaneous
domination of the leading and following sunspot fluxes in one of
the hemispheres results in the north--south asymmetry of the solar
magnetic flux. During the period of low activity of the Sun, the
imbalance is close to zero. It changes sign around the reversal of
the polar  magnetic field (Figure~\ref{leadfollow2}).

The sign of the imbalance of leading sunspots
(Figure~\ref{leadfollow2}) does not change during 11 years from
one reversal to the next and coincides with the sign of the polar
magnetic field in the northern hemisphere (shown at the top of
Figure~\ref{leadfollow2}). The sign of the imbalance of following
sunspots does not change either during the 11 years from one
reversal to the next and coincides with the sign of the polar
magnetic field in the southern hemisphere (shown at the bottom of
Figure~\ref{leadfollow2}).

\begin{figure}
\begin{center}
\includegraphics[width=0.95\textwidth]{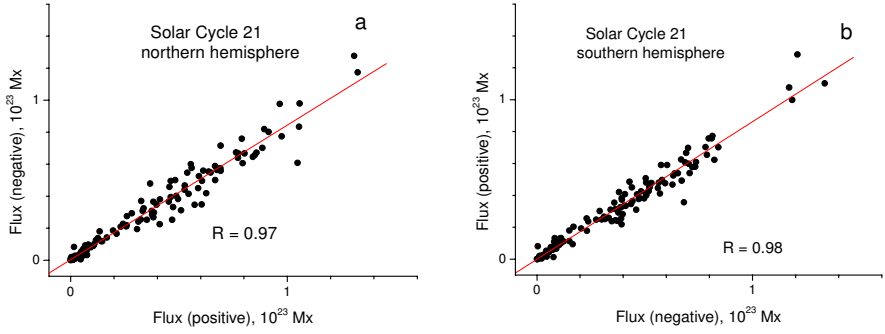}
   \caption{Correlation of positive and negative  magnetic fluxes
   (Solar Cycle 21): (a) northern hemisphere: $F_{{\rm N}}^{{\rm pos}}$ and $F_{{\rm N}}^{{\rm neg}}$;
   (b) southern hemisphere: $F_{{\rm S}}^{{\rm neg}}$ and $F_{{\rm S}}^{{\rm pos}}$.
   Regression line is shown in red. R: correlation coefficient.
   Each point represents one Carrington rotation.}
   \label{corrnorsou2}
\end{center}
\end{figure}

\begin{figure}[b]
\begin{center}
\includegraphics[width=3in]{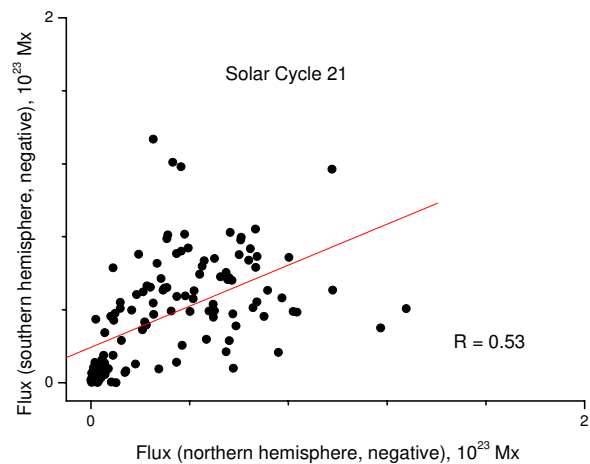}
   \caption{Correlation of negative magnetic fluxes of the northern
   [$F_{{\rm N}}^{{\rm neg}}$] and southern [$F_{{\rm S}}^{{\rm neg}}$]
   hemispheres (Solar Cycle 21).
   Regression line is shown in red. R: correlation coefficient.
   Each point represents one Carrington rotation.}
\label{corrns}
\end{center}
\end{figure}
The absolute values of the leading and following sunspot fluxes in
the same hemisphere are very well correlated. For Solar Cycle 21,
the flux of the following sunspots is plotted against the leading
sunspot fluxes in Figure~\ref{corrnorsou2}a (northern hemisphere)
and in Figure~\ref{corrnorsou2}b (southern hemisphere). The
correlation coefficients are 0.97 and 0.98, respectively. The
slope of the regression line is $0.84\pm 0.02$ for the northern
hemisphere (Figure~\ref{corrnorsou2}a) and $0.86\pm 0.02$ for the
southern one (Figure~\ref{corrnorsou2}b). The slopes for the two
hemispheres coincide within the uncertainties. The deviation of
this slope from unity is a measure of the leading--following
asymmetry of the flux concentration. Thus for the strong magnetic
fields [$B>100$\,G] the field-strength asymmetry between the
preceding and following polarities is nearly the same for the
northern and southern hemispheres. The correlation becomes much
poorer if one considers sunspots belonging to the opposite solar
hemispheres. For the same cycle, the following-sunspot fluxes
(southern hemisphere) are plotted against those of the
leading-sunspot fluxes (northern hemisphere) in
Figure~\ref{corrns}. The scatter of points is much greater, and
the correlation coefficient is only 0.53.

These observations are summarized in Table~\ref{correlation} where
correlation coefficients are given for different combinations of
the flux pairs (positive and negative fluxes, northern and
southern hemispheres). Correlation coefficients between the
leading and following sunspots that belong to the same hemisphere
show very high values. All correlation coefficients between the
fluxes of different hemispheres are much lower, varying from 0.52
to 0.58.

\begin{table}[h]
\caption{Correlation coefficients for magnetic fluxes of Solar
Cycle 21}
\label{correlation}       
\begin{tabular}{llll}
\hline\noalign{\smallskip}
Magnetic fluxes & $F_{{\rm N}}^{{\rm neg}}$ & $F_{{\rm S}}^{{\rm neg}}$ & $F_{{\rm S}}^{{\rm pos}}$  \\
\noalign{\smallskip}\hline\noalign{\smallskip}
$F_{{\rm N}}^{{\rm pos}}$ & 0.97 & 0.58 & 0.56 \smallskip \\
$F_{{\rm N}}^{{\rm neg}}$ &  & 0.53 & 0.52 \smallskip \\
$F_{{\rm S}}^{{\rm neg}}$ &  &  & 0.98 \\
\noalign{\smallskip}\hline
\end{tabular}
\end{table}

Domination of leading sunspots of one of the hemispheres over
leading sunspots of the other hemisphere is, therefore,
accompanied by domination of the following sunspots of the same
hemisphere. But the polarity of the leading and following sunspots
situated in the same hemisphere is opposite, therefore when we
take into account both types of sunspots (Figure~\ref{netflux2}),
we observe an imbalance lower than that of the leading sunspots
alone (Figure~\ref{leading}).

Thus, the observed effect of the magnetic-flux imbalance and its
change should be attributed to the major contribution of the
leading sunspot flux, which exceeds the contribution of the
following sunspots and determines the resulting sign of the
effect.

The following conclusions can be drawn on the basis of
Figures~\ref{netflux2}, \ref{leading}, and \ref{leadfollow2}:

i) The sign of the positive and negative flux imbalance remains
constant for 11 years from one reversal of the global magnetic
field to the next, coinciding with the sign of the  polar field in
the northern hemisphere (Figure~\ref{netflux2}).

ii) The sign of the imbalance is determined by the prevalence of
the leading-sunspot flux in one of the hemispheres over the
leading-sunspot flux of the opposite hemisphere
(Figure~\ref{leading}).

iii) If leading sunspots of a certain  hemisphere dominate leading
sunspots of the opposite hemisphere, then the following sunspots
of the first hemisphere will dominate the following sunspots of
the second one (Figure~\ref{leadfollow2}).

North--south asymmetry and imbalance of positive and negative
magnetic fluxes represent different, yet closely connected,
features of the photospheric magnetic field. The north--south
asymmetry (as defined by Equation~(\ref{deltans})) is the
difference of the total fluxes of the two solar hemispheres. In
contrast, the imbalance of positive and negative magnetic fluxes
represents the net flux of both hemispheres
(Equation~(\ref{deltaf})).

The regular change of the magnetic flux north--south asymmetry for
Solar Cycles $21-23$ described in Section~\ref{secnsasym} can be
deduced from conclusions i)\,--\,iii). We consider successively
the four quarters of a 22-year magnetic cycle of the Sun
(Table~\ref{polarity}).

For the first quarter (1976\,--\,1981), the imbalance was positive
[$\Delta F> 0$] (see Table~\ref{polarity} and
Figure~\ref{netflux2}) because of the major contribution  of
leading sunspots of the northern hemisphere, which have  a
positive sign during this period [$F_{{\rm N}}^{{\rm pos}} > 0$]
(Figure~\ref{leading}). Both positive and negative fluxes (leading
and following sunspots) of the northern hemisphere
(Figure~\ref{leadfollow2}) exceed the corresponding fluxes of the
leading and following sunspots of the opposite hemisphere
[$|F_{{\rm N}}^{{\rm pos}}|> |F_{{\rm S}}^{{\rm neg}}|$ and
$|F_{{\rm N}}^{{\rm neg}}|>|F_{{\rm S}}^{{\rm pos}}|$], which
means that the total magnetic flux of the northern hemisphere will
dominate.

For the second quarter (1982\,--\,1986),  the sign of the
imbalance changes [$\Delta F<0$]; now it is determined by leading
sunspots of negative sign, \textit{i.e.} by leading sunspots of
the southern hemisphere. Both the leading and following sunspots
of the southern hemisphere dominate the corresponding sunspots of
the northern hemisphere. As a result, after the reversal the total
flux of the southern hemisphere will dominate.

For the third quarter  (1987\,--\,1991),  the imbalance preserves
the negative sign, but after solar minimum, leading sunspots of
the negative sign appear in the northern hemisphere. Thus, a
negative sign of the imbalance points to the domination of leading
sunspots of the northern hemisphere. The magnetic fluxes of
leading (following) sunspots in the northern hemisphere exceed the
corresponding fluxes of the southern hemisphere. Now the total
flux of the northern hemisphere will dominate.

For the fourth quarter (1992\,--\,1996),  the sign of the
imbalance changes and becomes positive. It corresponds to the sign
of leading sunspots of the southern hemisphere. Domination of
leading (and following) sunspots of the southern hemisphere
results in the prevalence of the total magnetic flux of the
southern hemisphere.

In this way, the magnetic flux imbalance and the north--south
asymmetry for Solar Cycles $21-23$ proved to be linked.


\section{Conclusions}
   \label{secconcl}
We were concerned here with the photospheric magnetic fields
connected with solar active regions. Therefore, we confined
ourselves to the study of the near-equatorial helio-latitudes
($\pm 40^\circ$) and strong magnetic fields ($B>100$\,G). For the
northern and southern hemispheres, fluxes of positive and of
negative photospheric magnetic fields were evaluated to obtain
four resulting fluxes: $F_{{\rm N}}^{{\rm pos}}$, $F_{{\rm
N}}^{{\rm neg}}$, $F_{{\rm S}}^{{\rm pos}}$, and $F_{{\rm
S}}^{{\rm neg}}$. Since we considered the sunspot region and only
took into account strong fields, the main contribution to the flux
of positive (negative) polarity for each hemisphere could be
attributed to the flux of the leading or the following sunspots
with the corresponding polarities. Thus, each of the four fluxes
may be considered as the flux of leading (following) sunspots in
one of the hemispheres. The imbalance of the fluxes of positive
and negative polarities manifests itself in two forms: as the
imbalance of the fluxes within one hemisphere, and as the total
imbalance for both hemispheres.

i) {\it Imbalance of the fluxes for an individual hemisphere.} We
showed that for the leading and following sunspots of one
hemisphere, the moduli of their fluxes (for Solar Cycle 21) are
very strongly correlated (the correlation coefficient is 0.97 for
the northern hemisphere and 0.98 for the southern hemisphere), in
contrast to the fluxes of the sunspots from different hemispheres
(the correlation coefficient changes from 0.52 to 0.58). However,
the imbalance of the positive and negative fluxes of the solar
hemisphere shows significant (20\,\%) systematic changes during
the 22-year cycle. The sign of the difference between the positive
and negative fluxes is constant from one minimum to the next and
coincides with the sign of leading sunspots in the corresponding
hemisphere, which implies the domination of the leading-sunspot
flux over the flux of the following sunspots
(Figures~\ref{imbnorth2} and \ref{imbsouth2}). Thus, the sign of
imbalance is directly related  to the Hale-sign-law cycle.

ii) {\it Total imbalance.} For the whole interval of
helio-latitudes from $+40^\circ$  to $-40^\circ$ the difference
between the positive and the negative fluxes (4\,\% of the average
magnetic flux) also displays a 22-year periodicity; however, in
contrast to any separately considered hemisphere, the sign of the
total imbalance changes during the period of the Sun's global
magnetic-field reversal (Figure~\ref{netflux2}). The sign of the
imbalance does not change during the 11 years from one reversal to
the next and always coincides with the sign of the Sun's polar
magnetic field in the northern hemisphere. We showed that the
imbalance of the magnetic fluxes of the leading sunspots (15\,\%)
has the same sign as the total imbalance (Figure~\ref{leading})
and is in antiphase with the imbalance of the fluxes of the
following sunspots (Figure~\ref{leadfollow2}). Thus, the
domination of the leading sunspots in one of the hemispheres
determines the sign of the magnetic flux imbalance.

The sign of the imbalance between fluxes of leading sunspots of
two hemispheres  changes with a 22-year magnetic cycle in the same
way as the sign of the polar magnetic field in the northern
hemisphere, while the imbalance between fluxes of following
sunspots follows in sign the polar magnetic field in the southern
hemisphere (Figure~\ref{leadfollow2}).

The sign of the imbalance between fluxes of leading  sunspots of
the two hemispheres remains constant for each quarter of a 22-year
magnetic cycle (Figure~\ref{leadfollow2}). The imbalance of the
following sunspot fluxes is in antiphase with the imbalance of the
leading sunspot fluxes. This means that domination of leading
sunspots of a certain polarity is accompanied by the domination of
following sunspots with opposite polarity in the same hemisphere.

These features of the magnetic-flux imbalance can explain the
north--south asymmetry of the magnetic flux and its change in the
course of an 11-year solar cycle. The north--south asymmetry of
the magnetic flux displays a regular alternation of the dominant
hemisphere for  Solar Cycles $21-23$ (see Figure~\ref{nsasym} and
Table~\ref{polarity}). In the ascending phase (from solar-activity
minimum to the reversal) the northern hemisphere always dominates.
During the reversal the dominant role passes to the southern
hemisphere, which prevails until the next  minimum. Near the
solar-activity minimum, the northern hemisphere becomes dominant
again. Data considered in our analysis ($1976-2003$) refer to the
four Schwabe Cycles ($20-23$), when activity of the northern
hemisphere was leading in time (see Introduction). The regular
change of the north--south asymmetry of the magnetic flux that we
observed agrees with this effect.

The domination of the magnetic flux with a certain polarity displays
a 22-year periodicity; this polarity always coincides with the polar
magnetic-field polarity of the northern hemisphere.

The domination of leading sunspots in one of the hemispheres
determines the sign of the magnetic-flux imbalance. At the moment
of the reversal, the relative roles of leading sunspots change
with the subsequent domination of leading sunspots with opposite
polarity. During the solar-activity minimum, when the leading and
following sunspots change their signs according to Hale's law,
domination of leading sunspots of a certain polarity remains.
Before the minimum, leading sunspots of the southern hemisphere
dominate, while after the minimum the leading sunspots of the
northern hemisphere play the dominant role. Thus during the
22-year magnetic solar cycle we observe four periods with stable
configurations of magnetic fields, \textit{i.e.} periods
(quarters) of constant polarities of local and global magnetic
fields: from the minima, where the sunspot sign changes following
Hale's law, to the reversals of the Sun's global magnetic field,
and from the reversals to the minima.

We showed that both the sign of the north--south asymmetry of the
magnetic fluxes for Solar Cycles $21-23$ and the sign of the
imbalance of the positive and the negative fluxes are related to
one of  the quarters of the 22-year magnetic cycle where the
magnetic configuration of the Sun remains constant. Our results
enable us to express the change of the sign of the north--south
asymmetry and for the sign of the magnetic-flux imbalance simply
by Equations~(\ref{sign}) and (\ref{signf}). The sign of the
north--south asymmetry only depends on the phase of the 11-year
cycle  (before or after the reversal). On the other hand, the
imbalance sign depends both on the phase of the 11-year cycle and
on the parity of the solar cycle. It follows that the asymmetry of
the magnetic-field distribution develops according to a regular
pattern  and is closely related to the evolution of the local and
global magnetic fields.

\begin{acks}
NSO/Kitt Peak data used here are produced cooperatively by
NSF/NSO, NASA/GSFC, and NOAA/SEL. We thank
E.V. Miletsky for stimulating discussions and the referee for
many helpful comments.
\end{acks}




\end{article}

\end{document}